\documentclass[aps,notitlepage,onecolumn,nofootinbib,superscriptaddress,longbibliography]{revtex4-2}
 \usepackage[utf8]{inputenc}
\usepackage{times}
\usepackage{soul}
\usepackage{url}
\usepackage[utf8]{inputenc}
\usepackage[small]{caption}
\usepackage{graphicx}
\usepackage{amsmath}
\usepackage{amsthm}
\usepackage{booktabs}
\usepackage{algorithm}
\usepackage{algorithmic}
\usepackage{hyperref}
\hypersetup{colorlinks=true,citecolor=blue,linkcolor=blue,filecolor=blue,urlcolor=blue,breaklinks=true}
\usepackage{tikz}
\usetikzlibrary{positioning, shapes.geometric}

\usepackage{algorithm,algorithmic}
\usepackage{amssymb}
\usepackage{mathtools}
\usepackage{bm}
\usepackage{graphicx}
\usepackage{epsfig}
\usepackage{epstopdf}
\usepackage{subfigure}
\usepackage{array}
\usepackage{multirow}
\usepackage{booktabs}
\usepackage{eucal}
\usepackage{makecell}
\usepackage{fnbreak}

\usepackage{tikz} 

\usepackage[braket,qm]{qcircuit}
\usepackage{qcircuit}
\newcommand{\QSANN}{quantum self-attention neural network}

\newcommand{\lr}[1]{\left( #1 \right)}

\newcommand{\LR}[1]{\left\{ #1 \right\}}

\newcommand{\CC}{{{\mathbb C}}}

\newcommand{\ketbra}[2]{|#1\rangle\!\langle#2|}

\begin{document}
\title{Quantum Self-Attention Neural Networks for Text Classification}
\author{Guangxi Li}
\affiliation{Institute for Quantum Computing, Baidu Research, Beijing 100193, China}
\affiliation{Centre for Quantum Software and Information, University of Technology Sydney, NSW 2007, Australia}
\author{Xuanqiang Zhao}
\affiliation{Institute for Quantum Computing, Baidu Research, Beijing 100193, China}
\affiliation{QICI, Department of Computer Science, The University of Hong Kong, Hong Kong, China}
\author{Xin Wang}
\email{felixxinwang@hkust-gz.edu.cn}
\affiliation{Institute for Quantum Computing, Baidu Research, Beijing 100193, China}
\affiliation{Thrust of Artificial Intelligence, Information Hub, Hong Kong University of Science and Technology (Guangzhou), Nansha, China}

\begin{abstract}
An emerging direction of quantum computing is to establish meaningful quantum applications in various fields of artificial intelligence, including natural language processing (NLP). Although some efforts based on syntactic analysis have opened the door to research in Quantum NLP (QNLP), limitations such as heavy syntactic preprocessing and syntax-dependent network architecture make them impracticable on larger and real-world data sets. In this paper, we propose a new simple network architecture, called the quantum self-attention neural network (QSANN), which can compensate for these limitations. Specifically, we introduce the self-attention mechanism into quantum neural networks and then utilize a Gaussian projected quantum self-attention serving as a sensible quantum version of self-attention. As a result, QSANN is effective and scalable on larger data sets and has the desirable property of being implementable on near-term quantum devices. In particular, our QSANN outperforms the best existing QNLP model based on syntactic analysis as well as a simple classical self-attention neural network in numerical experiments of text classification tasks on public data sets. We further show that our method exhibits robustness to low-level quantum noises and showcases resilience to quantum neural network architectures.
\end{abstract}

\date{\today}
\maketitle

\section{Introduction}



\textit{Quantum computing} is a promising  paradigm \cite{preskill2021quantum} for fast computations that can provide substantial advantages in solving valuable problems \cite{harrow2017quantum,Childs2010,Montanaro2016,Childs2018a,Biamonte2017}. With major academic and industry efforts on developing quantum algorithms and quantum hardware, it has led to an increasing number of powerful applications in areas including optimization~\cite{Brandao2017}, cryptography~\cite{Xu2020}, chemistry~\cite{McArdle2018a,Cao2018b}, and machine learning~\cite{Biamonte2017,Rebentrost2014,huang2021power,Schuld2021}.

Quantum devices available currently known as the \textit{noisy intermediate-scale quantum} (NISQ) devices \cite{Preskill2018} have up to a few hundred physical qubits. They are affected by coherent and incoherent noise, making the practical implementation of many advantageous quantum algorithms less feasible. But such devices with 50-100 qubits already allow one to achieve quantum advantage against the most powerful classical supercomputers on certain carefully designed tasks \cite{arute2019quantum,zhong2020quantum}.
To explore practical applications with near-term quantum devices, plenty of NISQ algorithms \cite{bharti2021noisy,Cerezo2020,Endo2020} appear to be the best hope for obtaining quantum advantage in fields such as quantum chemistry \cite{Peruzzo2014}, optimization \cite{Farhi2014}, and machine learning \cite{Havlicek2019,Schuld2018,Mitarai2018,yang2022bert,qi2022classical,yang2023quantum}.
In particular, those algorithms 
dealing
with machine learning problems, by employing \textit{parameterized quantum circuits} (PQCs) \cite{Benedetti2019} (also called \textit{quantum neural networks} (QNNs) \cite{Farhi2018}), show great potential in the field of \textit{quantum machine learning} (QML).
See \cite{Yu2022_power,Caro2021,Li2022,Du2022,Jerbi2021,Yu2022,Huang2020a,Wang2020c,Zhao2021,Tian2023,Wang2023_ham,Abbas2021} for some recent progresses on the theory and applications of quantum neural networks in many directions.
However, in \textit{artificial intelligence} (AI), the study of QML in the NISQ era is still in its infancy. Thus it is desirable to 
explore more QML algorithms exploiting the power that lies within the NISQ devices.  

\textit{Natural language processing} (NLP) is a key subfield of AI that aims to give machines the ability to understand human language. Common NLP tasks include speech recognition, machine translation, text classification, etc., many of which have greatly facilitated our life. Due to human language's high complexity and flexibility, NLP tasks are generally challenging to implement. Thus, it is natural to think about whether and how quantum computing can enhance machines' performance on NLP. Some works focus on quantum-inspired language models \cite{Sordoni2013,Zhang2016b,Zhang2019d,basile2017towards} with borrowed ideas from quantum mechanics. Another approach, known as \textit{quantum natural language processing} (QNLP), seeks to develop quantum-native NLP models that can be implemented on quantum devices \cite{zeng2016quantum,meichanetzidis2020quantum,wiebe2019quantum,chen2020quantum}. Most of these QNLP proposals, though at the frontier, lack scalability as they are based on syntactic analysis, which is a preprocessing task requiring significant effort, especially for large data sets. Furthermore, these syntax-based methods employ different PQCs for sentences with different syntactical structures and thus are not flexible enough to process the innumerable complex expressions possible in human language. 

To overcome these drawbacks in current QNLP models, we propose the \textit{quantum self-attention neural network} (QSANN), where the self-attention mechanism is introduced into quantum neural networks. Our motivation comes from the excellent performance of self-attention on various NLP tasks such as language modeling \cite{devlin2018bert}, machine translation \cite{vaswani2017attention}, question answering \cite{li2019beyond}, and text classification \cite{guo2020multi}. We also note that a recently proposed method \cite{cha2020attention} for quantum state tomography, an important task in quantum computing, adopts the self-attention mechanism and achieves decent results.

In each quantum self-attention layer of QSANN, we first encode the inputs into high-dimensional quantum states, then apply PQCs on them according to the layout of the self-attention neural networks, and finally adopt a \textit{Gaussian projected quantum self-attention} (GPQSA) to obtain the output effectively.
To evaluate the performance of our model, we conduct numerical experiments of text classification with different data sets. The results show that QSANN outperforms the currently best known QNLP model as well as a simple classical self-attention neural network on test accuracy, implying potential quantum advantages of our method.
Our contributions are multi-fold:
\begin{itemize}
    \item Our proposal is the first QNLP algorithm with a detailed circuit implementation scheme based on the self-attention mechanism. This method can be implemented on NISQ devices and is more practicable on large data sets compared with previously known QNLP methods based on syntactic analysis.
    \item In QSANN, we introduce the Gaussian projected quantum self-attention, which can efficiently dig out the correlations between words in high-dimensional quantum feature space. Furthermore, visualization of self-attention coefficients on text classification tasks confirms its ability to focus on the most relevant words.
    \item We experimentally demonstrate that QSANN outperforms existing QNLP methods based on syntactic analysis \cite{lorenz2021qnlp} and simple classical self-attention neural networks on several public data sets for text classification. Numerical results also imply that QSANN is resilient to both quantum noises and quantum neural network architectures.
\end{itemize}

\subsection{Preliminaries and Notations}
\paragraph{Quantum Basics}
Here, some basic concepts about quantum computing necessary for this paper are briefly introduced (for more details, see \cite{Nielsen2002}).
In quantum computing, quantum information is usually represented by $n$-qubit (pure) quantum states over Hilbert space $\CC^{2^n}$. In particular, a pure quantum state could be represented by a unit vector $\ket{\psi} \in \CC^{2^n}$ (or $\bra{\psi}$), where the \textit{ket} notation $\ket{}$ denotes a column vector and the \textit{bra} notation $\bra{\psi}=\ket{\psi}^\dagger$ with $\dagger$ referring to conjugate transpose denotes a row vector.

The evolution of a pure quantum state $\ket{\psi}$ is mathematically described by applying a quantum circuit (or a quantum gate), i.e., $\ket{\psi^\prime}=U\ket{\psi}$, where $U$ is the unitary operator (matrix) representing the quantum circuit and $\ket{\psi^\prime}$ is the quantum state after evolution. Common single-qubit quantum gates include Hadamard gate $H$ and Pauli operators
\begin{align}
 H := \frac{1}{\sqrt{2}}\begin{bmatrix} 1 & 1\\ 1 & -1\end{bmatrix}, \
    X  := \begin{bmatrix} 0 & 1\\ 1 & 0\end{bmatrix}, 
    Y := \begin{bmatrix} 0 & -i\\ i & 0\end{bmatrix}, 
    Z := \begin{bmatrix} 1 & 0\\ 0 & -1\end{bmatrix},
\end{align}
and their corresponding rotation gates denoted by $R_P(\theta)$ $:= $ $ \text{exp}(-i\theta P/2)$ $ = \cos\frac{\theta}{2} I -i\sin\frac{\theta}{2}P$, where the rotation angle $\theta\in[0,2\pi)$ and $P\in\{X,Y,Z\}$. In this paper, multiple-qubit quantum gates mainly include the identity gate $I$, the CNOT gate and the tensor product of single-qubit gates, e.g., $Z\otimes Z$, $Z\otimes I$, $Z^{\otimes n}$ and so on.

Quantum measurement is a way to extract classical information from a quantum state. For instance, given a quantum state $\ket{\psi}$ and an observable $O$, one could design quantum measurements to obtain the information $\bra{\psi}O\ket{\psi}$. This work focuses on the hardware-efficient Pauli measurements, i.e., setting $O$ as Pauli operators or their tensor products. For instance, we could choose $Z_1 = Z\otimes I^{\otimes (n-1)}$, $X_2 =$ $I\otimes X\otimes I^{\otimes (n-2)}$, $Z_1Z_2 = Z\otimes Z\otimes I^{\otimes (n-2)}$, etc., with $n$ qubits in total.

\paragraph{Text Classification}
As one of the central and basic tasks in NLP field, text classification is to assign a given text sequence to one of the predefined categories. Examples of text classification tasks considered in this paper include topic classification and sentiment analysis. A commonly adopted approach in machine learning is to train a model with a set of pre-labeled sequences. When fed a new sequence, the trained model will be able to predict its category based on the experience learned from the training data set.

\paragraph{Self-Attention Mechanism}
In a self-attention neural network layer \cite{vaswani2017attention},
the input data $\{x_s\in \mathbb{R}^d \}_{s=1}^S$ are linearly mapped via three weight matrices, i.e., query $W_q\in \mathbb R^{d\times d}$, key $W_k\in \mathbb R^{d\times d}$ and value $W_v\in \mathbb R^{d\times d}$, to three parts $W_qx_s$, $W_kx_s$, $W_vx_s$, respectively, and by applying the inner product on the query and key parts, the output is computed as
\begin{align}
    y_s=\sum_{j=1}^S a_{s,j} \cdot W_vx_j \qquad \text{with} \qquad  a_{s,j}= \frac{\mathrm{e}^{x_s^{\top}W_q^{\top}W_kx_j}}{\sum_{l=1}^S\mathrm{e}^{x_s^{\top}W_q^{\top}W_kx_l}},
\end{align}
where $a_{s,j}$ denote the self-attention coefficients.

\section{Method}

In this section, we will introduce the QSANN in detail, which mainly consists of \textit{quantum self-attention layer} (QSAL), loss function, analytical gradients and analysis.

\begin{figure}[t]
	\centering
		\includegraphics[width=0.95\textwidth]{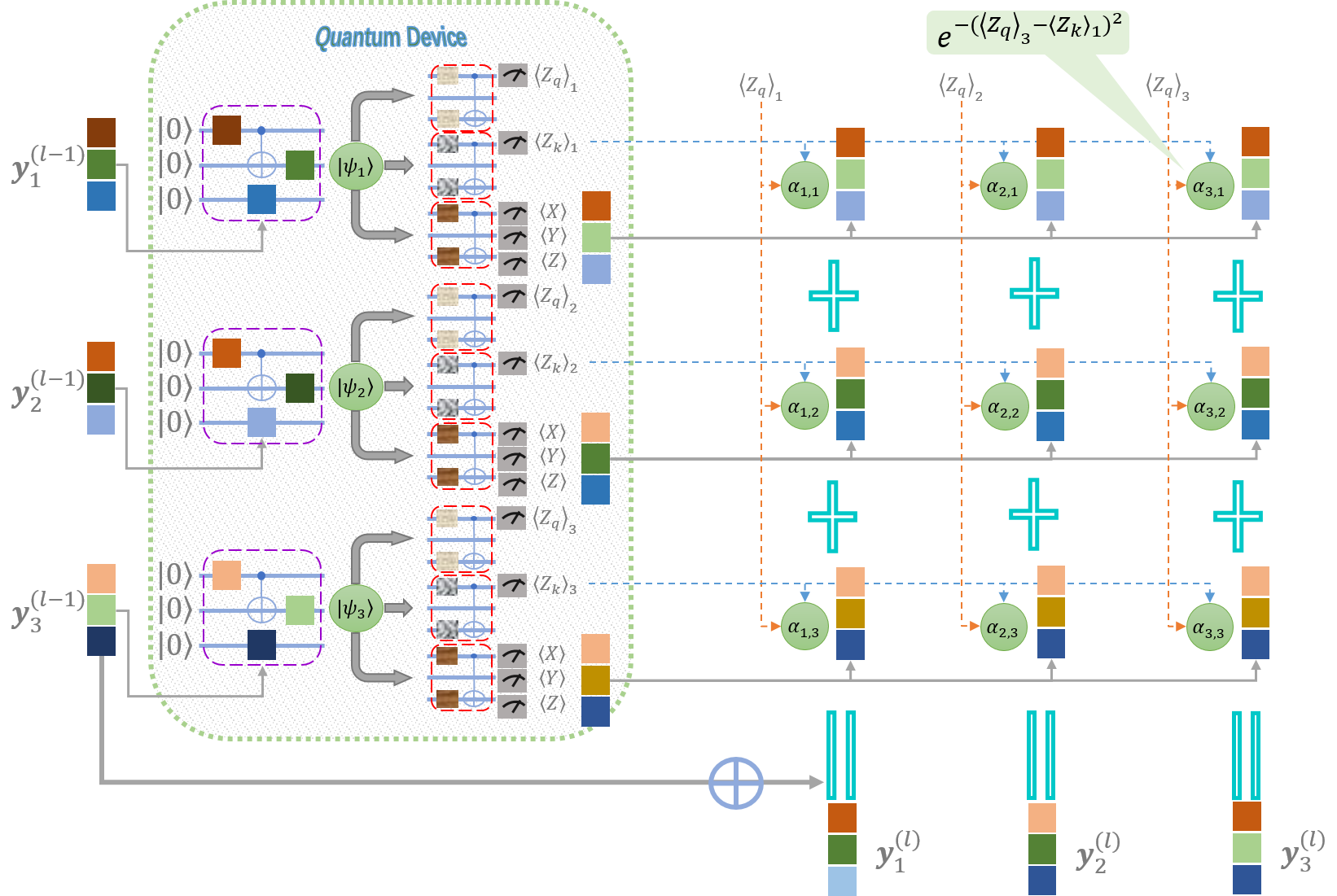}
	\caption{Sketch of a quantum self-attention layer (QSAL). On quantum devices, the classical inputs $\{\bm{y}^{(l-1)}_s\}$ are used as the rotation angles of quantum ansatzes (purple dashed boxes) to encode them into their corresponding quantum states $\{\ket{\psi_s}\}$.
    Then, a set of three ansatzes (in red dashed boxes) representing query, key, and value is applied to each state. Note that it is the same set of ansatzes applied to all the input states.
 On classical computers, the measurement outputs of the query part $\langle Z_q\rangle_s$ and the key part $\langle Z_k\rangle_j$ are computed through a Gaussian function to obtain the quantum self-attention coefficients $\alpha_{s,j}$ (green circles); we calculate classically weighted sums of the measurement outputs of the value part (small colored squares) and add the inputs to get the outputs $\{\bm y^{(l)}_s\}$, where the weights are the normalized coefficients $\Tilde{\alpha}_{s,j}$, cf. Eq. \eqref{eq:alpha}.}
	\label{fig:scheme_QSAL}
\end{figure}

\subsection{Quantum Self-Attention Layer}
In the classical self-attention mechanism \cite{vaswani2017attention}, there are mainly three components (vectors), i.e., queries, keys and values, where queries and keys are computed as weights assigned to corresponding values to obtain final outputs. 
Inspired by this mechanism, in QSAL we design the quantum analogs of these components. The overall picture of QSAL is illustrated in Fig. \ref{fig:scheme_QSAL}.

For the classical input data  $\{\bm{y}^{(l-1)}_s$ $\in \mathbb{R}^d \}$ of the $l$-th QSAL, we first use a quantum ansatz $U_{enc}$ to encode them into an $n$-qubit quantum Hilbert space, i.e.,
\begin{align}\label{eq:input_encoding}
    \ket{\psi_s}=U_{enc}(\bm{y}^{(l-1)}_s)H^{\otimes n}\ket{0^n}, \quad 1\le s \le S,
\end{align}
where $H$ denotes the Hadamard gate and $S$ denotes the number of input vectors in a data sample.

Then we use another three quantum ansatzes, i.e., $U_q$, $U_k$, $U_v$ with parameters $\bm\theta_q$, $\bm\theta_k$, $\bm\theta_v$, to represent the query, key and value parts, respectively.
Concretely, for each input state $\ket{\psi_s}$, we denote by $\langle Z_q\rangle_s$ and $\langle Z_k\rangle_s$ the Pauli-$Z_1$ measurement outputs of the query and key parts, respectively, where 
   \begin{align}  \label{eq:query_key}
    \langle Z_q\rangle_s & := \bra{\psi_s} U_q^{\dagger}(\bm\theta_q) Z_1  U_q(\bm\theta_q) \ket{\psi_s},  \nonumber \\ 
    \langle Z_k\rangle_s & := \bra{\psi_s} U_k^{\dagger}(\bm\theta_k) Z_1  U_k(\bm\theta_k) \ket{\psi_s}.
\end{align} 
The measurement outputs of the value part are represented by a $d$-dimensional vector
\begin{align}\label{eq:value}
    \bm o_s :=\begin{bmatrix}
    \langle P_1\rangle_s &
    \langle P_2\rangle_s &
    \cdots &
    \langle P_d\rangle_s
    \end{bmatrix}^\top,
\end{align}
where $\langle P_j\rangle_s$ $= \bra{\psi_s} U_v^{\dagger}(\bm\theta_v) P_j  U_v(\bm\theta_v) \ket{\psi_s}$.
Here, each $P_j\in\LR{I,X,Y,Z}^{\otimes n}$ denotes a Pauli observable.

Finally, by combining Eqs. \eqref{eq:query_key} and \eqref{eq:value}, the classical output  $\{\bm y^{(l)}_s\in \mathbb{R}^d \}$ of the $l$-th QSAL are computed as follows:
\begin{align}\label{eq:qsann_output}
    \bm y^{(l)}_s= \bm{y}^{(l-1)}_s + \sum_{j=1}^S \Tilde{\alpha}_{s,j} \cdot\bm o_j,
\end{align}
where $\Tilde{\alpha}_{s,j}$ denotes the normalized quantum self-attention coefficient between the $s$-th and the $j$-th input vectors and is calculated by the corresponding query and key parts:
\begin{align}\label{eq:alpha}
    \Tilde{\alpha}_{s,j}=\frac{\alpha_{s,j}}{\sum_{m=1}^S \alpha_{s,m}} \qquad \text{with} \qquad  \alpha_{s,j} :=\text{e}^{-( \langle Z_q\rangle_s -  \langle Z_k\rangle_j)^2}.
\end{align}
Here in Eq. \eqref{eq:qsann_output}, we adopt a residual scheme when computing the output, which is analogous to  \cite{vaswani2017attention}.

\paragraph{Gaussian Projected Quantum Self-Attention}
\label{sec:GPQSA}
When designing a quantum version of self-attention, a natural and direct extension of the inner-product self-attention to consider is ${\alpha}_{s,j}:= |\bra{\psi_s} U_q^{\dagger} U_k \ket{\psi_j}|^2$.
However, due to the unitary nature of quantum circuits, $\bra{\psi_s} U_q^{\dagger} U_k$ can be regarded as rotating $\ket{\psi_s}$ by an angle, which makes it difficult for $\ket{\psi_s}$ to simultaneously correlate those $\ket{\psi_j}$ that are far away. In a word, this direct extension is not suitable or reasonable for working as the quantum self-attention.
Instead, the particular quantum self-attention proposed in Eq. \eqref{eq:alpha}, which we call \textit{Gaussian projected quantum self-attention} (GPQSA), could overcome the above drawback. In GPQSA, the states $U_q \ket{\psi_s}$ (and $U_k \ket{\psi_j}$) in large quantum Hilbert space are projected to classical representations $\langle Z_q\rangle_s$ (and $\langle Z_k\rangle_j$) in one-dimensional\footnote{Multi-dimension is also possible by choosing multiple measurement results, like the value part.}
classical space via quantum measurements and a Gaussian function is applied to these classical representations.
As $U_q$ and $U_k$ are separated, it is pretty easy to correlate $\ket{\psi_s}$ to any $\ket{\psi_j}$, making GPQSA more suitable to serve as a quantum self-attention.
Here, we utilize the Gaussian function \cite{micchelli2006universal,yang2023quantum} mainly because it contains infinite-dimensional feature space and is well-studied in classical machine learning. 
Numerical experiments also verify our choice of Gaussian function. 
We also note that other choices for building quantum self-attention are also worth future study.

\textbf{Remark.}  During the preparation of this manuscript,
we became aware that Ref.~\cite{di2022dawn} also made initial attempts to employ the attention mechanism in QNNs. In that work, the authors mentioned a possible quantum extension towards a quantum Transformer where the straightforward inner-product self-attention is adopted. As discussed above, the inner-product self-attention may not be reasonable for dealing with quantum data. In this work, we present that GPQSA is more suitable for the quantum version of self-attention and show the validity of our method via numerical experiments on several public data sets.


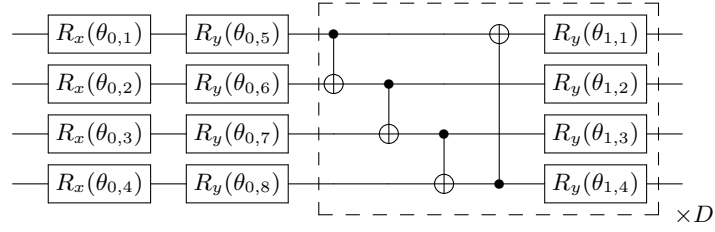
\begin{figure}[t]
\centering
\[\Qcircuit @C=1.5em @R=0.5em {
  & \gate{R_x(\theta_{0,1})} & \gate{R_y(\theta_{0,5})}  &\ctrl{+1}& \qw & \qw  &\targ &  \gate{R_y(\theta_{1,1})}&\qw 
  \\ 
  & \gate{R_x(\theta_{0,2})} & \gate{R_y(\theta_{0,6})}  & \targ & \ctrl{+1} &\qw&\qw &  \gate{R_y(\theta_{1,2})} & \qw 
  \\ 
  & \gate{R_x(\theta_{0,3})} & \gate{R_y(\theta_{0,7})}  & \qw &\targ& \ctrl{+1} &\qw&  \gate{R_y(\theta_{1,3})} & \qw 
  \\ 
  & \gate{R_x(\theta_{0,4})} & \gate{R_y(\theta_{0,8})}  & \qw&\qw &\targ & \ctrl{-3} &  \gate{R_y(\theta_{1,4})} & \qw 
  \gategroup{1}{4}{4}{8}{1.0em}{--} \\
  &&&&&&&& \ \ \ \ \times D 
}\]
\caption{The ansatz used in QSANN. The first two columns denote the $R_x$-$R_y$ rotations on each single-qubit subspace, followed by repeated CNOT gates  and single-qubit $R_y$ rotations. The block circuit in the dashed box is repeated $D$ times to enhance the expressive power of the ansatz.}
\label{fig:ansatz}
\end{figure}

\paragraph{Ansatz Selection}
In QSAL, we employ multiple ansatzes for the various components, i.e., data encoding, query, key and value. Hence, we give a brief review of it here.

In general, an ansatz, a.k.a. parameterized quantum circuit \cite{Benedetti2019}, has the form
$
    U({\bm \theta})=\prod_{j} U_j(\theta_j)V_j,
$
where $ U_j(\theta_j)=  \exp(-i\theta_j P_j/2)$ and $V_j$ denotes a fixed operator such as Identity, CNOT and so on. Here, $P_j$ denotes a Pauli operator. Due to the numerous choices of the form of $V_j$, various kinds of ansatzes can be used. In this paper, we use the strongly entangled ansatz \cite{Schuld2018} shown in Fig. \ref{fig:ansatz} in QSAL. This circuit has $n(D+2)$ parameters in total for $n$ qubits and $D$ repeated layers.

\begin{figure*}[t]
	\centering
		\includegraphics[width=0.8\textwidth]{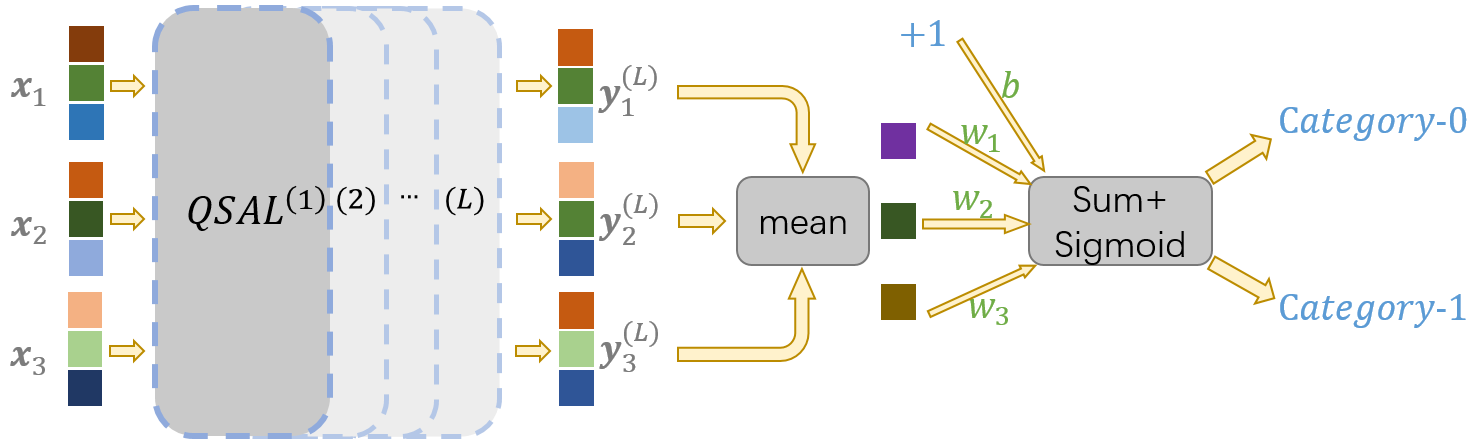}
	\caption{Sketch of QSANN, where a sequence of classical vectors $\{\bm x_s\}$ firstly goes through $L$ QSALs to obtain the corresponding sequence of feature vectors $\{\bm y_s^{(L)}\}$, then through the average operation, and finally through the fully-connected layer for the binary prediction task.}
	\label{fig:scheme_qsann}
\end{figure*}

\subsection{Loss Function}


Consider the data set $\mathcal{D}$ $:=$ $\{({\bm x_{m;1}}$, ${\bm x_{m;2}}$, $\ldots$, ${\bm x_{m;S_m}} )$, ${y_m}\}_{m=1}^{N_s}$, where 
    there are in total $N_s$ sequences or samples and each has $S_m$ words with a label ${y_m}$ $\in\LR{0,1}$.
Here, we assume each word is embedded as a $d$-dimensional vector, i.e., ${\bm x_{m;s}}\in \mathbb R^d$.
The whole procedure of QSANN is depicted in Fig. \ref{fig:scheme_qsann}, which mainly consists of $L$ QSALs to extract hidden features and one fully-connected layer to complete the binary prediction task.
Here, the mean squared error \cite{Ziegel1999} is employed as the loss function:
\begin{align}\label{eq:loss}
    \mathcal L\lr{\bm \Theta,\bm w,b;\mathcal{D}}=\frac{1}{2N_s} \sum_{m=1}^{N_s} \lr{{\hat y}_{m}-{y}_{m} }^2 + \text{RegTerm},
\end{align}
where the predicted value ${\hat y}_{m}$ is defined as
    ${\hat y}_{m}:=\sigma\lr{\bm w^\top \cdot \frac{1}{S_m} \sum_{s=1}^{S_m} {\bm y}^{(L)}_{m;s} +b}$
with $\bm w\in \mathbb R^d$ and $b\in \mathbb R$ denoting the weights and bias of the final fully-connected layer, $\bm \Theta$ denoting all parameters in the ansatz, $\sigma$ denoting the sigmoid activation function and `RegTerm' being the regularization term to avoid overfitting in the training process.

Combining Eqs. \eqref{eq:input_encoding}
- \eqref{eq:alpha}, we know each output of QSAL is dependent on all its inputs, i.e.,
\begin{align}
    {\bm y}^{(l)}_{m;s} :=& {\bm y}^{(l)}_{m;s} \lr{\bm \theta^{(l)}_q,\bm \theta^{(l)}_k,\bm \theta^{(l)}_v; \{{\bm y}^{(l-1)}_{m;i}\}_{i=1}^{S_m}} \nonumber \\
    =& {\bm y}^{(l-1)}_{m;s} + \sum_{j=1}^{S_m} \Tilde{\alpha}_{s,j}^{(l)} \lr{\bm \theta^{(l)}_q,\bm \theta^{(l)}_k; \{{\bm y}^{(l-1)}_{m;i}\}_{i=1}^{S_m}}
    \cdot \bm o_j^{(l)} \lr{\bm \theta^{(l)}_v; {\bm y}^{(l-1)}_{m;j}},
\end{align}
where ${\bm y}^{(0)}_{m;s}= {\bm x}_{m;s}$ and $1\le s \le S_m$, $1\le l\le L$. Here, the regularization term is defined as
\begin{align}
   \text{RegTerm} := \frac{\lambda}{2d} \|\bm w\|^2 + \frac{\gamma}{2d}  \sum_{s=1}^{S_m}\|{\bm x}_{m;s}\|^2,
\end{align}
where $\lambda,\gamma \ge 0$ are two regularization coefficients. 

With the loss function defined in Eq. \eqref{eq:loss}, we can optimize its parameters by (stochastic) gradient-descent \cite{Bottou2004}.
The analytical gradient analysis can be found in Sec. \ref{sec:gradient}. Finally, with the above preparation, we could train our QSANN to get the optimal (or sub-optimal) parameters. See Algorithm \ref{alg:qsann_text_classification}
for details on the training procedure. We remark that if the loss converges during training or the maximum number of iterations is reached, the optimization stops.

\subsection{Analytical Gradients}
\label{sec:gradient}
Here, we give the stochastic analytical partial gradients of the loss function with regard to its parameters as follows.
We first consider the parameters in the last \QSANN{} layer, i.e., $\bm \theta^{(L)}_q, \bm \theta^{(L)}_k, \bm \theta^{(L)}_v$, and the final fully-connected layer, i.e., $\bm w,b$. Then the parameters in the front layers could be evaluated similarly and be updated through the back-propagation algorithm \cite{Goodfellow-et-al-2016}. Given the $m$-th data sample $\LR{\lr{\bm x_1, \bm x_2, \ldots,\bm  x_{S_m} },y}$ (here, we omit $m$ in the subscript for writing convenience, the same below), we have 
\begin{align}
    &\frac{\partial \mathcal L}{\partial \bm w} = \Tilde{\sigma} \cdot \frac{1}{S_m} \sum_{s=1}^{S_m} \bm{y}^{(L)}_s+\frac{\lambda}{d}\bm w, \qquad
    \frac{\partial \mathcal L}{\partial b}  = \Tilde{\sigma}, \\
    &\frac{\partial \mathcal L}{\partial \bm{y}^{(L)}_s}  = \Tilde{\sigma} \cdot \frac{1}{S_m}\cdot \bm w,
\end{align}
where $\Tilde{\sigma}=\lr{\sigma-y}\cdot \sigma\lr{1-\sigma}$ and $\sigma$ denotes the abbreviation of $\sigma\lr{\bm w^\top \cdot \frac{1}{S_m} \sum_{s=1}^{S_m}{\bm y}^{(L)}_s +b}$.
We also have
\begin{small}
\begin{align}
     \frac{\partial \mathcal L}{\partial \bm \theta^{(L)}_v} &= \sum_{s=1}^{S_m} \lr{\frac{\partial \mathcal L}{\partial \bm{y}^{(L)}_s}}^\top \sum_{j=1}^{S_m} \frac{\partial \bm{y}^{(L)}_s}{\partial \bm o_j^{(L)}} \cdot \frac{\partial \bm o_j^{(L)}}{\partial \bm \theta^{(L)}_v}, \label{eq:partial_derivative_v}
     \\
     \frac{\partial \mathcal L}{\partial \bm \theta^{(L)}_q} &= \sum_{s=1}^{S_m} \lr{\frac{\partial \mathcal L}{\partial \bm{y}^{(L)}_s}}^\top \sum_{j=1}^{S_m} \frac{\partial \bm{y}^{(L)}_s}{\partial \alpha_{s,j}^{(L)}} \cdot \frac{\partial \alpha_{s,j}^{(L)}}{\partial \langle Z_q\rangle_s}\cdot \frac{\partial \langle Z_q\rangle_s}{\partial \bm \theta^{(L)}_q},  \label{eq:partial_derivative_q}
     \\  \label{eq:partial_derivative_k}
      \frac{\partial \mathcal L}{\partial \bm \theta^{(L)}_k} &= \sum_{s=1}^{S_m} \lr{\frac{\partial \mathcal L}{\partial \bm{y}^{(L)}_s}}^\top \sum_{j=1}^{S_m} \frac{\partial \bm{y}^{(L)}_s}{\partial \alpha_{s,j}^{(L)}} \cdot \sum_{i=1}^{S_m} \frac{\partial \alpha_{s,j}^{(L)}}{\partial \langle Z_k\rangle_i}\cdot \frac{\partial \langle Z_k\rangle_i}{\partial \bm \theta^{(L)}_k},
\end{align}
\end{small}
where ${\partial \bm{y}^{(L)}_s}/{\partial \bm o_j^{(L)}}$  $= \alpha_{s,j}^{(L)}$, ${\partial \bm{y}^{(L)}_s}/{\partial \alpha_{s,j}^{(L)}}$  $=\bm o_j^{(L)}$, ${\partial \alpha_{s,j}^{(L)}}/{\partial \langle Z_q\rangle_s}=$  $-\sum_{i=1}^{S_m} {\partial \alpha_{s,j}^{(L)}}/{\partial \langle Z_k\rangle_i}$  and 
\begin{align}
    \frac{\partial \alpha_{s,j}^{(L)}}{\partial \langle Z_k\rangle_i} =  - \alpha_{s,j}^{(L)}&\lr{ \alpha_{s,i}^{(L)}-  \delta_{ij}}\cdot 2\lr{ \langle Z_q\rangle_s- \langle Z_k\rangle_i},   \nonumber \\
    & \delta_{ij}=
\begin{cases}
1,& i=j\\
0,& \text{otherwise}.
\end{cases}
\end{align}
Furthermore, the last three partial derivatives of Eqs. \eqref{eq:partial_derivative_v}, \eqref{eq:partial_derivative_q} and \eqref{eq:partial_derivative_k} could be evaluated directly on the quantum computers via the parameter shift rule \cite{Mitarai2018}. For example, 
\begin{align}
    \frac{\partial \langle Z_q\rangle_s}{\partial \theta^{(L)}_{q,j}} = \frac{1}{2}\lr{\langle Z_q\rangle_{s,+}- \langle Z_q\rangle_{s,-}},
\end{align}
where $\langle Z_q\rangle_{s,\pm} :=\bra{\psi_s}U_{q,\pm}^{\dagger} Z U_{q,\pm} \ket{\psi_s}$ and $ U_{q,\pm}$ $:=$ $U_q\lr{\bm\theta_{q,-j}^{(L)},\theta_{q,j}^{(L)}\pm \frac{\pi}{2}}$.

Finally, in order to derive the partial derivatives of the parameters in the front layers, we also need the following:
\begin{small}
\begin{align}
    \frac{\partial \mathcal L}{\partial \bm{y}^{(L-1)}_i} &= \frac{\partial \mathcal L}{\partial \bm{y}^{(L)}_i} + \sum_{s=1}^{S_m} \lr{\frac{\partial \mathcal L}{\partial \bm{y}^{(L)}_s}}^\top  \frac{\partial \bm{y}^{(L)}_s}{\partial \bm o_i^{(L)}} \cdot \frac{\partial \bm o_i^{(L)}}{\partial  \bm{y}^{(L-1)}_i}  \nonumber\\
   & + \lr{\frac{\partial \mathcal L}{\partial \bm{y}^{(L)}_i}}^\top \sum_{j=1}^{S_m}  \frac{\partial \bm{y}^{(L)}_i}{\partial \alpha_{i,j}^{(L)}} \cdot \frac{\partial \alpha_{i,j}^{(L)}}{\partial \langle Z_q\rangle_i}\cdot \frac{\partial \langle Z_q\rangle_i}{\partial \bm{y}^{(L-1)}_i} \nonumber \\
  &  +  \sum_{s=1}^{S_m} \lr{\frac{\partial \mathcal L}{\partial \bm{y}^{(L)}_s}}^\top \sum_{j=1}^{S_m}  \frac{\partial \bm{y}^{(L)}_s}{\partial \alpha_{s,j}^{(L)}} \cdot \frac{\partial \alpha_{s,j}^{(L)}}{\partial \langle Z_k\rangle_i}\cdot \frac{\partial \langle Z_k\rangle_i}{\partial \bm{y}^{(L-1)}_i},
\end{align}
\end{small}
where the four terms denote the residual, value, query and key parts, respectively, and each sub-term can be evaluated similarly to the above analysis.
With the above preparation, we could easily calculate every parameter's gradient and update these parameters accordingly.

\renewcommand{\algorithmicrequire}{\textbf{Input:}}
\renewcommand{\algorithmicensure}{\textbf{Output:}}
\begin{algorithm}[H]
\caption{QSANN training for text classification}
\label{alg:qsann_text_classification}
\begin{algorithmic} 
\REQUIRE The training data set $\mathcal{D}$ $:=$ $\{({\bm x_{m;1}}$, ${\bm x_{m;2}}$, $\ldots$, ${\bm x_{m;S_m}} )$, ${y}_m\}_{m=1}^{N_s}$, $EPOCH$, number of QSALs $L$ and optimization procedure 
\ENSURE The final ansatz parameters $\Theta^*$, weight $\bm{w}^*, b^*$ 
\STATE Initialize the ansatz parameters $\Theta$, weight $\bm w$ from Gaussian distribution $\mathcal N(0,0.01)$ and the bias $b$ to 0.
\FOR{$ep=1,\dots,EPOCH$}
\FOR{$m=1,\ldots,N_s$}
\STATE Apply the encoder ansatz $U_{enc}$ to each of ${\bm x_{m;s}}$ to get the corresponding quantum state $\ket{\psi_s}$, cf. \eqref{eq:input_encoding}.
\STATE Apply $U_q$ and $U_k$ to $\ket{\psi_s}$ and measure the Pauli-Z expectations to get $\langle Z_q\rangle_s,\langle Z_k\rangle_s$, cf. \eqref{eq:query_key}, and then calculate the quantum self-attention coefficients $\alpha_{s,j}$, cf. \eqref{eq:alpha}.
\STATE Apply $U_v$ and measure a series of Pauli expectations to get $\bm o_s$, cf. \eqref{eq:value}, and then compute the output $\{\bm y_s^{(l)}\}$ of the $l$-th QSAL, cf. \eqref{eq:qsann_output}.
\STATE Repeat 4-6 $L$ times to get the output $\{\bm y_s^{(L)}\}$ of the $L$-th QSAL.
\STATE Average $\{\bm y_s^{(L)}\}$ and through a fully-connected layer to obtain the predicted value ${\hat y}_{m}$.
\STATE Calculate the mean squared error in \eqref{eq:loss} and update the parameters through the optimization procedure.
\ENDFOR
\IF{the stopping condition is met}
\STATE Break.
\ENDIF
\ENDFOR
\end{algorithmic}
\end{algorithm}

\subsection{Analysis of QSANN}

According to the definition of the Quantum Self-Attention Layer, for a sequence with $S$ words, we need $S(d+2)$ Pauli measurements to obtain the $d$-dimensional value vectors as well as the queries and keys for all words from the quantum device. After that, we need to compute $S^2$ self-attention coefficients for all $S^2$ pairs of words on the classical computer. In general, QSANN takes advantage of quantum devices' efficiency in processing high-dimensional data while outsourcing some calculations to classical computers. This approach keeps the quantum circuit depth low and thus makes QSANN robust to low-level noise common in near-term quantum devices. This beneficial attribute is further verified by numerical results in the next section, where we test QSANN against noise.

In short, our QSANN first encodes words into a large quantum Hilbert space as the feature space and then projects them back to low-dimensional classical feature space by quantum measurement. Recent works have proved rigorous quantum advantages on some classification tasks by utilizing high-dimensional quantum feature space \cite{liu2021rigorous} and projected quantum models \cite{huang2021power}. Thus, we expect that our QSANN might also have the potential advantage of digging out some hidden features that are classically intractable.
{Furthermore, the low-parameter variational quantum circuit exhibits the ability to achieve low generalization error \cite{qi2023theoretical} with few training data \cite{caro2022generalization}, providing further evidence for the effectiveness of our QSANN method.}
In the following section, we carry out numerical simulations of QSANN on several data sets to evaluate its performance on binary text classification tasks.

\section{Numerical Results}

In order to demonstrate the performance of our proposed QSANN, we have conducted numerical experiments on public data sets, where the quantum part was accomplished via classical simulation. Concretely, we first exhibit the better performance of QSANN by comparing it with i) the 
syntactic analysis-based quantum model \cite{lorenz2021qnlp} on two simple tasks, i.e., MC and RP, ii) the \textit{classical self-attention neural network} (CSANN) and the naive method on three public sentiment analysis data sets, i.e., Yelp, IMDb and Amazon \cite{kotzias2015group}. 
Then we show the reasonableness of our particular quantum self-attention GPQSA via visualization of self-attention coefficients. Next, we perform noisy experiments to show the robustness of QSANN to noisy quantum channels. 
Finally, we perform noisy experiments with different ansatzes to demonstrate the resilience of QSANN to the architectures of quantum neural networks.
All the simulations and optimization loops are implemented via Paddle Quantum\footnote{https://github.com/paddlepaddle/Quantum} on the PaddlePaddle Deep Learning Platform~\cite{Ma2019p}.

\paragraph{Data Sets}
The two simple synthetic data sets we employed come directly from \cite{lorenz2021qnlp}, which are named MC and RP, respectively.
MC contains 17 words and 130 sentences (70 train + 30 development + 30 test) with 3 or 4 words each; RP has 115 words and 105 sentences (74 train + 31 test) with 4 words in each one.
The other three data sets we use are real-world data sets available at \cite{Dua:2019} as the Sentiment Labelled Sentences Data Set. 
These data sets consist of reviews of restaurants, movies and products selected from Yelp, IMDb and Amazon, respectively. Each of the three data sets contains 1000 sequences, where half are labeled as `0' (for negative) and the other half as `1' (for positive). And each sequence contains several to dozens of words. We randomly select 80\% as training sequences and the rest 20\% as test ones.


\begin{table}[t]
\centering
\caption{Overview of hyper-parameter settings. Here, `LR' denotes learning rate, $D_{enc}, D_q, D_k, D_v$ denote the depths of the corresponding ansatzes and $d=n(D_{enc}+2)$.}
\begin{tabular}{cccccccc}
\toprule
Data set & $n$ & $d$ & $D_{enc}$ & $D_{q/k/v}$ & $\lambda$ & $\gamma$ & LR    \\ \midrule
MC      & 2 & 6 &    1      &1      &    0     &    0    &  0.008     \\ \midrule
RP      &4 & 24  &   4   &5             &     0.2    &    0.4    &    0.008   \\ \midrule
Yelp    & 4 &12  & 1& 1           & 0.2     & 0.2    & 0.008 \\ \midrule
IMDb    & 4& 12  &    1     &1        &     0.002    &   0.002     &   0.002    \\ \midrule
Amazon  & 4& 12  &      1   &2           &     0.2    &    0.2    & 0.008       \\ \toprule
\end{tabular}
\label{table:hyper_params_setting}
\end{table}

\label{sec:limitation}
\paragraph{Experimental Setting}
In the experiments, we use a single self-attention layer for both QSANN and CSANN. As a comparison, we also perform the most straightforward method, i.e., directly averaging the embedded vectors of a sequence, followed by a fully-connected layer, which we call the `Naive' method, on the three data sets of reviews. Here, we note that only comparing these simple classical models is because there are still significant restrictions on current quantum hardware. It is pretty unfair to compare with the most potent classical models.

\textbf{Remark.}
We note that due to the current limitations of quantum hardware, using mini- or small-scale tasks for benchmarking is a common practice in current QNLP research. 
Additionally, the quantum transformer is still in its infancy, and it may not be fair to directly compare it with the most advanced classical transformers or hybrid transformers \cite{yang2022bert}
currently available. Despite all this, we believe QSANN provides a good starting point for demonstrating the potential advantages and applications of quantum computing in NLP, providing valuable experience and insights for more in-depth research in the future.

In QSANN, all the encoder, query, key and value ansatzes have the same qubit number and are constructed according to Fig. \ref{fig:ansatz}, which are easily implementable on the NISQ devices.
Specifically, assuming the $n$-qubit encoder ansatz has $D_{enc}$ layers with $n(D_{enc}+2)$ parameters, we just set the dimension of the input vectors as $d=n(D_{enc}+2)$. The depths of the query, key and value ansatzes are set to the same and are, at most, the polynomial size of the qubit number $n$. 
The actual hyper-parameter settings on different data sets are concluded in Table \ref{table:hyper_params_setting}. 
In addition, we choose $Z_1,\ldots, Z_n$, $X_1,\ldots, X_n$, $Y_1,\ldots, Y_n$ as the Pauli observables $P_j$ in Eq. \eqref{eq:value}. For example, it is just required $3n$ observables when $D_{enc}=1$. However, if $D_{enc}>1$, we could also choose two-qubit observables $Z_{12}, Z_{23}$ and so on.
All the ansatz parameters $\bm\Theta$ and weight $\bm w$ are initialized from a Gaussian distribution with zero mean and 0.01 standard deviation, and the bias $b$ is initialized to zero.
Here, the ansatz parameters are not initialized uniformly from $[0,2\pi)$ is mainly due to the residual scheme applied in Eq. \eqref{eq:qsann_output}. 
During the optimization iteration, we use Adam optimizer \cite{Kingma2014}. And we repeat each experiment 9 times with different parameter initializations to collect the average accuracy and the corresponding fluctuations.

In CSANN, we set $d=16$ and the classical query, key and value matrices are also initialized from a Gaussian distribution with zero mean and 0.01 standard deviation. Except for these, almost all other parameters are set the same as QSANN. These settings and initializations are the same in the naive method as well.

\begin{table}[t]
\centering
\caption{Training accuracy and test accuracy of QSANN as well as DisCoCat on MC and RP tasks.}
\begin{tabular}{ccccccc}
\toprule
\multirow{2}{*}{Method} & \multicolumn{3}{c}{MC}          & \multicolumn{3}{c}{RP}         \\ \cmidrule(r){2-4} \cmidrule(r){5-7} 
                        & \# Paras & TrainAcc(\%) & TestAcc(\%) & \# Paras & TrainAcc(\%) & TestAcc(\%) \\ \midrule
DisCoCat \cite{lorenz2021qnlp}               &    40      &     83.10      &    79.80      &     168     &       90.60    &     \textbf{72.30}     \\ \midrule
QSANN                   &   25       &    \textbf{100.00}       &   \textbf{100.00}       &     109     &  \textbf{95.35$\pm$1.95} & 67.74$\pm$0.00 \\ \toprule
\end{tabular} 
\label{table:res_MC_RP}
\end{table}

\paragraph{Results on MC and RP Tasks}
The results on MC and RP tasks are summarized in Table \ref{table:res_MC_RP}. In the MC task, our method QSANN could easily achieve a 100\% test accuracy while requiring only 25 parameters (18 in query-key-value part and 7 in fully-connected part). However, in DisCoCat, the authors use 40 parameters but get a test accuracy lower than 80\%. This result strongly demonstrates the powerful ability of QSANN for binary text classification. Here, the parameters in the encoder part are not counted as they could be replaced by fixed representations such as pre-trained word embeddings.
In the RP task, we get a higher training accuracy but a slightly lower test accuracy. However, we observe that both test accuracies are pretty low when compared with the training accuracy. It is mainly because there is a massive bias between the training set and test set, i.e., more than half of the words in the test set have not appeared in the training one. Hence, the test accuracy highly depends on random guessing.

\begin{table*}[t]
\centering
\caption{Test accuracy of QSANN compared to CSANN and the naive method on Yelp, IMDb, and Amazon data sets. The highest accuracy in each column is indicated in bold font. On all the three data sets, QSANN achieves the highest accuracies among the three methods while using much fewer parameters than CSANN.}
\begin{tabular}{ccccccc}
\toprule
\multirow{2}{*}{Method} & \multicolumn{2}{c}{Yelp} & \multicolumn{2}{c}{IMDb} & \multicolumn{2}{c}{Amazon} \\ \cmidrule(r){2-3} \cmidrule(r){4-5}\cmidrule(r){6-7} 
                        & \# Paras    & TestAcc (\%)  & \# Paras    & TestAcc (\%)  & \# Paras     & TestAcc  (\%)  \\ \midrule
Naive   &  17 &  82.78$\pm$0.78  &   17   &  79.33$\pm$0.67  &  17   &   80.39$\pm$0.61          \\ \midrule
CSANN  & 785  & 83.11$\pm$0.89   &  785 &  79.67$\pm$0.83  & 785  &    83.22$\pm$1.28   \\ \midrule
QSANN   & 49   &  \textbf{84.79$\pm$1.29}  &  49   &   \textbf{80.28$\pm$1.78}  &  61  & \textbf{84.25$\pm$1.75}   \\ \toprule
\end{tabular} 
\label{table:res_yelp_imdb_amazon}
\end{table*}

\paragraph{Results on Yelp, IMDb and Amazon Data Sets}
As there are no quantum algorithms for text classification on these three data sets before, we benchmark our QSANN with the classical self-attention neural network (CSANN). The naive method is also listed for comparison.
The results on Yelp, IMDb and Amazon data sets are summarized in Table \ref{table:res_yelp_imdb_amazon}. We can intuitively see that QSANN outperforms CSANN and the naive method on all three data sets. Specifically, CSANN has 785 parameters (768 in classical query-key-value part and 17 in fully-connected part) on all data sets. In comparison, QSANN has only 49 parameters (36 in query-key-value part and 13 in fully-connected part) on the Yelp and IMDb data sets and 61 parameters (48 in query-key-value part and 13 in fully-connected part) on the Amazon data set, improving the test accuracy by about 1\% as well as saving more than 10 times the number of parameters.
Therefore, QSANN could have a potential advantage for text classification.

\begin{figure}[t]
	\centering
		\includegraphics[width=0.6\textwidth]{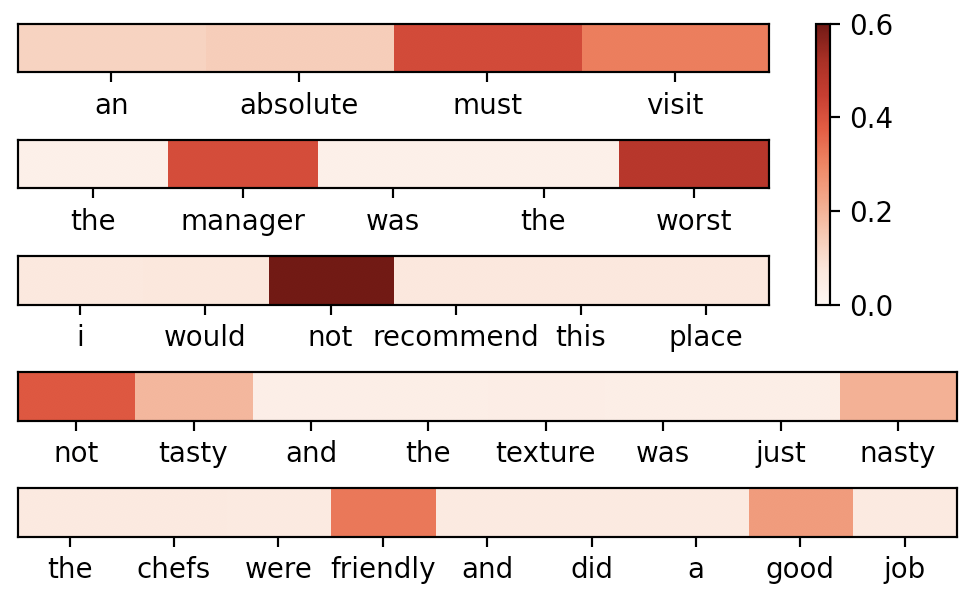}
	\caption{Heat maps of the averaged quantum self-attention coefficients for some selected test sequences from the Yelp data set, where a deeper color indicates a higher coefficient. Words that are more sentiment-related are generally assigned higher self-attention coefficients by our Gaussian projected quantum self-attention, implying the validity and interpretability of QSANN.}
	\label{fig:qsa_visual}
\end{figure}

\paragraph{Visualization of Self-Attention Coefficient}
To intuitively demonstrate the reasonableness of the Gaussian projected quantum self-attention, in Fig.~\ref{fig:qsa_visual} we visualize the averaged quantum self-attention coefficients of some selected test sequences from the Yelp data set. Concretely, for a sequence, we calculate $\frac{1}{S}\sum_{s=1}^S\Tilde{\alpha}_{s,j}$ for $j=1,\ldots,S$ and visualize them via a heat map, where $S$ is the number of words in this sequence and $\Tilde{\alpha}_{s,j}$ is the quantum self-attention coefficient. As shown in the figure, words with higher quantum self-attention coefficients are indeed those that determine the emotion of a sequence, implying the power of QSANN for capturing the most relevant words in a sequence on text classification tasks.


\begin{figure}[t]
	\centering
	\subfigure[]{\label{fig:noise_ansatz}
		\includegraphics[width=0.45\textwidth]{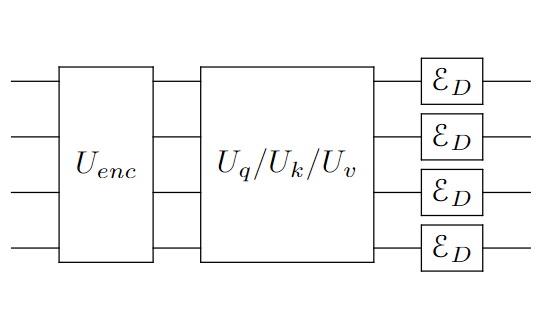}}
	\subfigure[]{\label{fig:noise_yelp}
		\includegraphics[width=0.45\textwidth]{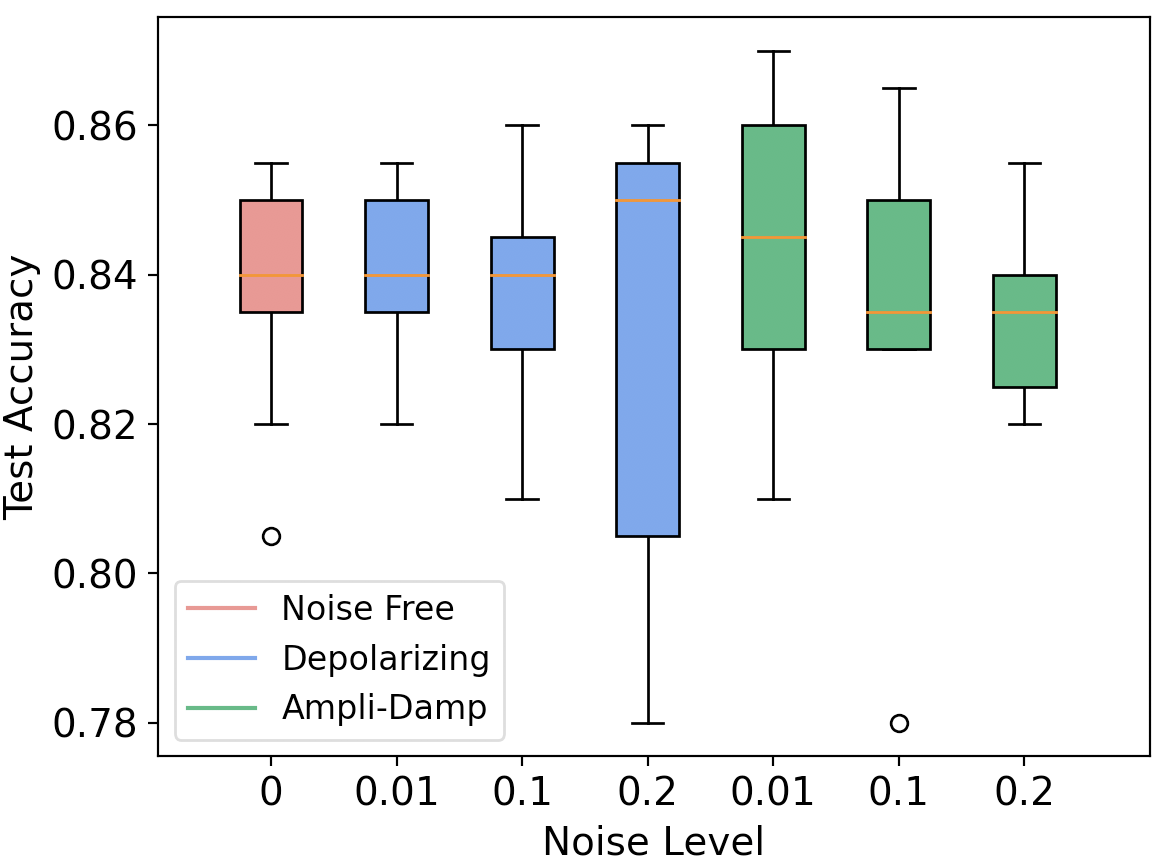}}
	\caption{(a) The diagram for adding depolarizing channels in our simulated experiments. The amplitude-damping channels are added in the same way. (b) Box plots of test accuracy on Yelp data set with depolarizing and amplitude damping noises. 
Each box contains nine repeated experiments.
 The absence of a notable decrease in accuracy implies the noise-resilience attribute of QSANN.}
\end{figure}

\paragraph{Noisy Experimental Results on Yelp Data Set}
\label{sec:noise_exp}
Due to the limitations of the near-term quantum computers, we add experiments with noisy quantum circuits to demonstrate the robustness of QSANN on the Yelp data set. We consider the representative channels~\cite{Nielsen2002} such as the depolarizing channel 
$\mathcal{E}_D(\rho)$
and the amplitude-damping channel
$\mathcal{E}_{AD}(\rho)$:
\begin{align} \label{eq:depolarizing}
    \mathcal{E}_D(\rho) &:=\lr{1-p}\rho + \frac{p}{3}\lr{ X\rho X + Y\rho Y +Z\rho Z}, \\
  \mathcal{E}_{AD}(\rho) & := E_0 \rho E_0^\dagger + E_1 \rho E_1^\dagger,
\end{align}
with
$E_0=\ketbra{0}{0}+\sqrt{1-p}\ketbra{1}{1}$ and $E_1=\sqrt{p}\ketbra{0}{1}$ denoting the Kraus operators.
Here, $\rho=\ketbra{\psi}{\psi}$ for a pure quantum state $\ket{\psi}$ and $p$ denotes the noise level.
As a regular way to analyze the effect of quantum noises, we add these single-qubit noisy channels in the final circuit layer to represent the whole system's noise, which is illustrated in Fig. \ref{fig:noise_ansatz}.


We take the noise level $p$ as 0.01, 0.1, 0.2  for these two noisy channels, respectively, and the box plots of test accuracies are depicted in Fig. \ref{fig:noise_yelp}. From the picture, we see the test accuracy of our QSANN almost does not decrease when the noise level is less than 0.1, 
and even when the noise level is up to 0.2, the overall test accuracy has only decreased a little, showing that QSANN is robust to these quantum noises.


\begin{figure}[t]
\centering
\subfigure[Ansatz-0]{
\Qcircuit @C=1.0em @R=0.5em {
   & \gate{R_x} & \gate{R_y}  &\ctrl{+1}& \qw & \qw  &\targ &  \gate{R_y}&\qw 
  \\ 
  & \gate{R_x} & \gate{R_y}  & \targ & \ctrl{+1} &\qw&\qw &  \gate{R_y} & \qw 
  \\ 
  & \gate{R_x} & \gate{R_y}  & \qw &\targ& \ctrl{+1} &\qw&  \gate{R_y} & \qw 
  \\ 
  & \gate{R_x} & \gate{R_y}  & \qw&\qw &\targ & \ctrl{-3} &  \gate{R_y} & \qw 
  \gategroup{1}{4}{4}{8}{1.0em}{--} \\
  &&&&&&&& \ \ \  \times D 
}} \qquad\qquad
\subfigure[Ansatz-1]{
\Qcircuit @C=1.5em @R=0.5em {
 & \gate{R_x} & \gate{R_y}  &\ctrl{+1}& \qw & \ctrl{+3}  & \gate{R_y}&\qw 
  \\ 
  & \gate{R_x} & \gate{R_y}  & \targ & \ctrl{+1} &\qw&  \gate{R_y} & \qw 
  \\ 
  & \gate{R_x} & \gate{R_y}  & \ctrl{+1} &\targ& \qw &  \gate{R_y} & \qw 
  \\ 
  & \gate{R_x} & \gate{R_y}  & \targ  &\qw &\targ  &  \gate{R_y} & \qw 
  \gategroup{1}{4}{4}{7}{1.0em}{--} \\
  &&&&&&& \ \ \  \times D 
}} 
\subfigure[Ansatz-2]{
\Qcircuit @C=1.0em @R=0.5em {
  & \gate{R_x} & \gate{R_y}  &\ctrl{+1}& \qw & \qw   &  \gate{R_y}&\qw 
  \\ 
  & \gate{R_x} & \gate{R_y}  & \targ & \ctrl{+1} &\qw&  \gate{R_y} & \qw 
  \\ 
  & \gate{R_x} & \gate{R_y}  & \qw &\targ& \ctrl{+1} & \gate{R_y} & \qw 
  \\ 
  & \gate{R_x} & \gate{R_y}  & \qw&\qw &\targ &  \gate{R_y} & \qw 
  \gategroup{1}{4}{4}{7}{1.0em}{--} \\
  &&&&&&& \ \ \  \times D 
}} \qquad\qquad
\subfigure[Ansatz-3]{
\Qcircuit @C=0.8em @R=0.5em {
  & \gate{R_x} & \gate{R_y}  &\ctrl{+1}& \ctrl{+2}& \ctrl{+3}  &\qw & \qw &\qw & \gate{R_y}&\qw 
  \\ 
  & \gate{R_x} & \gate{R_y}  & \targ & \qw &\qw& \ctrl{+1}& \ctrl{+2} & \qw & \gate{R_y} & \qw 
  \\ 
  & \gate{R_x} & \gate{R_y}  & \qw &\targ& \qw &\targ   & \qw &\ctrl{+1}& \gate{R_y} & \qw 
  \\ 
  & \gate{R_x} & \gate{R_y}  & \qw&\qw &\targ & \qw  & \targ & \targ  &  \gate{R_y} & \qw 
  \gategroup{1}{4}{4}{10}{1.0em}{--} \\ 
  &&&&&&&&&& \ \ \  \times D 
}}
\caption{Four types of ansatzes used in QSANN. Each has a different entangled layer.}
\label{fig:4ansatzes}
\end{figure}
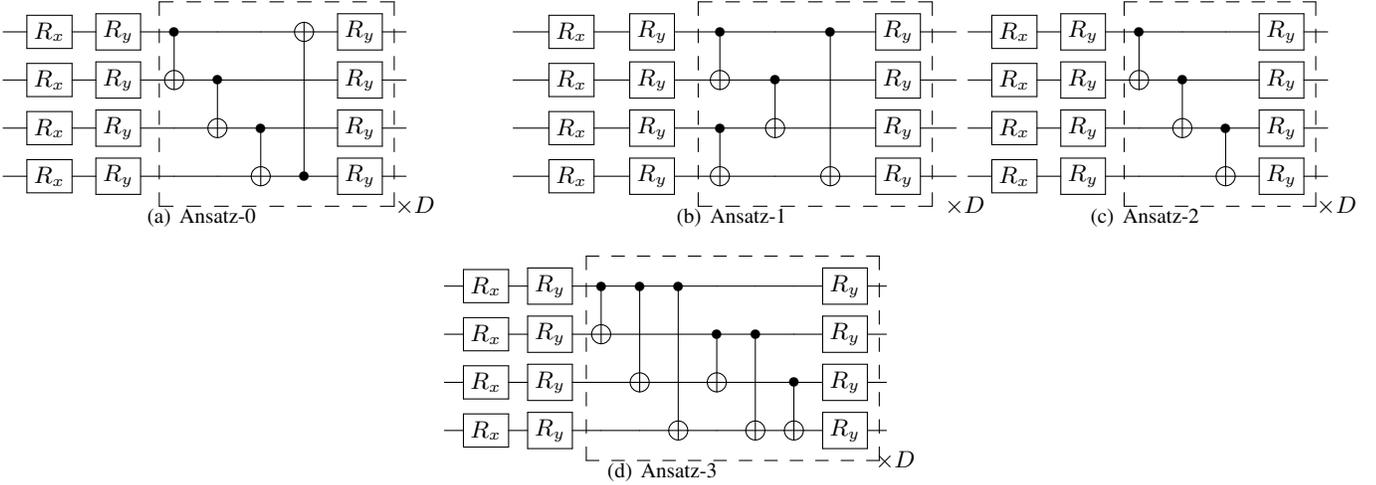

\begin{table}[t]
\centering
\begin{tabular}{ccccccc}
\toprule
\multirow{2}{*}{Method} & \multicolumn{2}{c}{MC}          & \multicolumn{2}{c}{RP}         \\ \cmidrule(r){2-3} \cmidrule(r){4-5} 
                         & TrainAcc(\%) & TestAcc(\%)  & TrainAcc(\%) & TestAcc(\%) \\ \midrule
Ansatz-0           &      100.00      &    100.00      &           94.74$\pm$1.20   &     67.74$\pm$0.00    \\ \midrule
Ansatz-1           &      100.00      &    100.00      &           94.71$\pm$0.11   &     67.03$\pm$0.72    \\ \midrule
Ansatz-2           &      100.00      &    100.00      &           94.71$\pm$0.93  &     67.38$\pm$0.36   \\ \midrule
Ansatz-3              &    100.00    &   100.00    &  94.74$\pm$0.15 & 67.74$\pm$0.00 \\ \toprule
\end{tabular} 
\caption{Training accuracy and test accuracy of QSANN with four different ansatzes on MC and RP tasks.}
\label{table:res_MC_RP_4ansatzes}
\end{table}

\paragraph{Noisy Experimental Results with Different Ansatzes}

Given the recent limitations of quantum hardware topology, some ansatzes are easier to implement than others. As such, exploring the performance of QSANN under different ansatzes is crucial to determining the difficulty level in deploying QSANN on current quantum hardware. Additionally, it is worth investigating which ansatz can most easily achieve optimal performance of QSANN for specific practical tasks.

In this subsection, we test QSANN using different ansatzes on both MC and RP data sets. As depicted in Fig. \ref{fig:4ansatzes}, these ansatzes utilize different entanglement layers while keeping the single-qubit gates and the total number of parameters unchanged. Furthermore, a depolarizing channel with $p=0.1$ is added to each ansatz, as shown in Eq. \eqref{eq:depolarizing}. Other settings remain the same as in Table \ref{table:hyper_params_setting}. The final results are shown in Table \ref{table:res_MC_RP_4ansatzes}, where we see that the performance of the four ansatz types is virtually identical. This directly indicates that QSANN is resilient to ansatz architectures.

\section{Discussions}
\label{sec:conclusion}

We have proposed a quantum self-attention neural network (QSANN) by introducing the self-attention mechanism to quantum neural networks. Specifically, the adopted Gaussian projected quantum self-attention exploits the exponentially large quantum Hilbert space as the quantum feature space, making QSANN have the potential advantage of mining some hidden correlations between words that are difficult to dig out classically. Numerical results show that QSANN outperforms the best-known QNLP method and a simple classical self-attention neural network for text classification on several public data sets. Moreover, using only shallow quantum circuits and Pauli measurements, QSANN can be easily implemented on near-term quantum devices and is noise-resilient, as implied by simulation results. We believe that this attempt to combine self-attention and quantum neural networks would open up new avenues for QNLP as well as QML. 

As a future direction, more advanced techniques such as positional encoding and multi-head attention can be employed in quantum neural networks for generative models and other more complicated tasks. 
Another exciting future research direction is to move toward large language models. However, we must realize that there are still many challenges and limitations to overcome, particularly in the NISQ era. Despite these challenges, our work represents a promising step towards this goal, and we are optimistic about the potential of quantum computing in NLP. As quantum hardware continues to evolve and improve, we anticipate that our methods can be gradually extended to more complex algorithms and tasks, unlocking new possibilities for QNLP research.






\textbf{Acknowledgements.} 
We would like to thank Prof. Sanjiang Li and Prof. Yuan Feng for their helpful discussions.
We also thank Zihe Wang and Chenghong Zhu for their help related to the experiments.
G. L. acknowledges the support from the Baidu-UTS AI Meets Quantum project, the China Scholarship Council (No. 201806070139), and the Australian Research Council project (Grant No: DP180100691).
Part of this work was done when X. Z. and X. W. were at Baidu Research.



\bibliographystyle{unsrt}
\bibliography{qsann}

\end{document}